\documentclass[twocolumn,aps,prb]{revtex4}   % style for Physical Review B and AJP are similar
\usepackage{amsmath,amssymb}    % need for subequations
\usepackage{graphicx}   % for figures
\usepackage{dcolumn}    % for tables

\begin{document}

\title{Two falling-chain demonstrations based on Einstein's equivalence 
principle }
\author{Chun Wa Wong}
\email{cwong@physics.ucla.edu}   %optional
\author{Seo Ho Youn}
\author{Kosuke Yasui}
 \affiliation{Department of Physics and Astronomy, 
University of California, Los Angeles, CA 90095-1547} 
\date{September 22, 2006} 

\begin{abstract}
Simple demonstrations based on the equivalence principle are given of how a folded 
chain and a horizontal flat chain fall down when one chain end is fixed to a rigid 
support.
\end{abstract}

\maketitle

We describe here how Einstein's equivalence principle can be used to show how the 
falling ends of two flexible chains fall when one chain end in each chain is held fixed 
to a rigid support. The first of these falling chains is a folded chain that has its 
two ends initially close together, with an initial horizontal separation $\Delta x = 0$. 
Kucharski\cite{Kucharski41} has given a one-dimensional continuum model of this 
falling chain. He finds from its Lagrangian equation of motion that its energy is 
conserved and is concentrated in the falling arm. The falling arm is thus forced to 
fall faster than $g$, the acceleration due to gravity. The falling time turns out 
to be 85\% of that for free fall, making it relatively straightforward to confirm the 
theoretical description.\cite{footnote} A description of this falling chain based 
directly on energy conservation can be found in many places,\cite{Wong06} including
Ref.~\onlinecite{Hamel49,Calkin89,Thornton04}. 

Calkin and March\cite{Calkin89} have obtained experimental confirmation of energy 
conservation of the falling folded chain by measuring its tension at the 
support as a function of the vertical falling distance. They show that their 
experimental result can be reproduced closely by the Kucharski one-dimensional 
continuum model, except near the end of fall when the last link in the chain 
turns over as a rigid body.\cite{Wong06} The recent history of this falling chain 
problem has been reviewed by Wong and Yasui\cite{Wong06} who point out that energy 
conservation arises from the fact that links are transferred to the stationary arm 
at the bend of the flexible chain by elastic collisions, not inelastic ones. The 
falling chain end falls faster than $g$ because that part of the chain immediately 
next to the falling end is below it and pulls it down further. Wong and Yasui are 
able to isolate the downward chain tension involved with the help of the Lagrangian 
formulation of mechanics. 

Recently, Tomaszewski, Pieranski and Geminard\cite{TPG06} have re-confirmed the faster 
than $g$ fall of the folded chain both by experiment and by numerical simulation. 
They have also extended their study to falling chains with nonzero initial separations 
$\Delta x$. For the limiting case of $\Delta x \approx L$ (the chain length), where the 
chain is initially stretched taut as an almost flat catenary, they find that the 
chain end falls (almost) freely on release, with an acceleration very close to $g$. 
Their video capture pictures show clearly that the falling end is part of a freely 
falling horizontal chain segment that shrinks in length as the chain falls. 

A simple way to ``see'' that the falling end is falling freely is to note that the 
chain near the falling end is (almost) horizontal so that there is little mass next to 
and below the falling end to drag it down further, and no mass next to and above 
the falling end to pull it up. In other words, the falling end falls freely because it 
is part of a freely falling horizontal chain segment. The reason why the horizontal 
chain segment itself falls freely is because the chain is flexible and bends 
smoothly and without kink to the original horizontal shape towards the falling end. 
This bend starts from the supported end of the chain and travels outward towards the 
falling end as the chain falls. The chain tension begins at the supported end and 
follows the direction of the chain itself, becoming horizontal beyond the bend. It can 
thus only pull the horizontal chain segment towards the supported end without affecting 
the horizontal segment's acceleration in the vertical direction. Finally, the smoothness 
of the chain at the bend is the result of Newton's third law that states that action 
and reaction forces must be equal in strength and opposite in direction. This is possible 
only when the chain has no kink. If there were a kink, the tangent direction of the chain 
would change discontinuously across the kink to give a reaction force that is not 
opposite in direction to the action, in violation of Newton's third law.

Historically, the nature of these falling chains have been clarified by using high-speed 
photography,\cite{Schagerl97,Taft} sensor and electronic circuitry,\cite{Calkin89} and 
video capture\cite{TPG06}. We shall demonstrate the motion of these falling chains using 
only a chain and a smooth horizontal surface provided by a desk or table, with the result 
interpreted by the equivalence principle of Einstein.   
  
The principle of mass equivalence, proposed by Einstein in 1907,\cite{Einstein07} 
states that the gravitational and inertial masses of an object are always equal. 
This means that the acceleration acquired by a freely falling observer in a 
gravitational field is just the acceleration $g$ due to gravity. A freely falling 
observer then sees no gravity.\cite{Einstein20} The sudden realization of this 
connection between inertia and weight was for Einstein the ``happiest thought'' 
of his life.\cite{Einstein20} 

\begin{figure}
\includegraphics{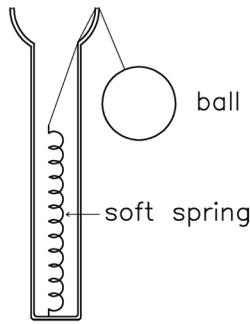}
\caption{\label{fig:toy}
The cup, soft spring and hanging metal ball in the toy made by Eric Rogers for 
Einstein's 76th birthday. } 
\end{figure}

The weightlessness of a freely falling observer is beautifully demonstrated by a 
toy made by Eric Rogers of Princeton University and given to Einstein as a puzzle 
on the occasion of Einstein's 76th and last birthday in 1955.\cite{Rogers79} A cup 
with a hollow tube at its bottom, looking like a champagne glass with a long stem, is
placed at the end of a broomstick. A long soft spring is attached to the inside base 
of the hollow tube. The soft spring is stretched at the free end by a long thread that 
connects it to a metal ball outside the cup so that the ball hangs over the lip of 
the cup, as shown schematically in Fig.~\ref{fig:toy}. The problem is to find a 
{\it foolproof} way to move the hanging metal ball at the end of the long thread 
into the cup.

The old man solved the problem ``at once''.\cite{Rogers79} Holding the toy 
high by the end of the broomstick, he allowed the toy to fall freely along 
the length of the long broomstick. The outside mass became weightless. The 
soft spring then retracted, pulling the ball into the cup, to the old man's 
great delight.\cite{Cohen79} This story is behind the title ``An Old Man's Toy'' 
of Zee's book on gravity.\cite{Zee89}

\begin{figure}
\includegraphics{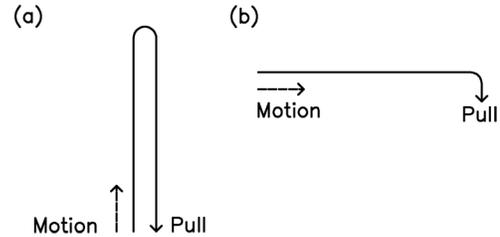}
\caption{\label{fig:chains}
Pulling on chains placed on a horizontal surface: (a) the folded chain, and (b) the 
straight chain. }
\end{figure}

Let us now return to our falling folded chain. If its falling end falls freely, 
it will remain at rest relative to a freely falling observer. We now propose to show 
what this freely falling observer actually sees by putting a light folded ball or
beaded chain on a smooth horizontal table top made of glass, marble, steel, 
Formica or polished wood where the friction is likely to be relatively small. Hold  
one chain end by the hand above the table. This chain end represents the original chain 
end at the fixed support. It is next made to move relative to the freely falling observer 
by being pulled suddenly away from the fold of the chain in a horizontal
direction along the length of the folded chain, as shown in Fig.~\ref{fig:chains}(a). 
You the observer will see that the free end still on the table will not remain at rest 
on the table, as it should if it 
were falling freely in a falling folded chain. You will find instead that it will move 
in the direction {\it opposite} to the pulling direction, thereby showing that the 
free end of the falling folded chain is pulled by the chain next to and below it to 
fall faster than $g$. The pull and the motion to be observed are sketched in the figure.
Marking the initial position of the chain end on the table with a small object placed 
next to it will help to define its subsequent motion. We recommend using about a meter 
length of a light ball chain like those used to operate overhead fans. (It is sometimes 
called a No. 6 ball chain in the United States). These light chains will not scratch 
your table if you do it only a few times. 

The demonstration of a falling flat chain is just as easy. Arrange the ball chain 
on the table in a straight line. Lift the chain by the end representing the support, 
and pull it suddenly in a horizontal direction {\it perpendicular} to the length of 
the chain on the table, as shown in Fig.~\ref{fig:chains}(b). You will see that the 
chain on the table is pulled along its 
length towards the bend of the chain, but it will not move in the direction of the 
pull or opposite to it, thereby showing that the falling end of the original chain  
falls freely. In both demonstrations, the equivalence principle allows the observer 
to effectively fall freely in space so that the relative motion of the falling chain 
end can be seen more clearly by the unaided eye.

These demontrations are most effective when the pull at the ``support'' 
end is of middling strength, fast enough to make friction unimportant but 
slow enough to make the motion to be demonstrated clearly seen
by the eye. Of course, our pull is unlikely to generate a constant 
acceleration or a value necessarily close to $g$. The demonstrations 
work even with variable accelerations of any convenient strength. This 
is indeed one of their charms. 

The falling flat chain can also be demonstrated directly but perhaps not as persuasively 
in the following way: A demonstrator stretches the chain horizontally before letting one 
end fall down. The observer should stand some 
distance away from the demonstrator in order to follow the fall readily with the eye. 
It is possible to see a falling horizontal chain segment before it finally merges into 
the rotating arm at the support end. One can also see that the kinetic energy of the 
rotating arm is supplied by the energy carried in by the chain links transferred from 
the falling horizontal chain segment.  However, the observer cannot tell for sure that 
the falling horizontal chain segment is falling freely.


\begin{thebibliography}{5}
\bibitem{footnote} Experimental confirmation of Kucharski's observation in early years 
has been mentioned in other papers, and we remember seeing in an old book on mechanics 
a photograph of a falling folded chain falling faster than a small freely falling ball. 
However, we are not able to locate any specific reference on these observations.
\end{thebibliography}
\end{document}